\documentstyle[12pt]{article}
  \advance\oddsidemargin  by -1in
  \advance\evensidemargin by -1in
  \advance\topmargin      by -0.5in
\def\lsim{\mathrel{\rlap{\lower4pt\hbox{\hskip1pt$\sim$}}
    \raise1pt\hbox{$<$}}}         
\def\gsim{\mathrel{\rlap{\lower4pt\hbox{\hskip1pt$\sim$}}
    \raise1pt\hbox{$>$}}}         

\def\beq{\begin{equation}}
\def\endeq{\end{equation}}
\def\arr{\begin{eqnarray}}
\def\endarr{\end{eqnarray}}
\makeindex

\begin{document}

\begin{center}
{\bf \huge
Asymptotic behaviour \\
of the total cross section of
$p-p$ scattering\\
and the Akeno cosmic ray data
\vspace{0.5cm}\\}
{\Large
N.N.Nikolaev
\medskip\\}
{\sl IKP(Theorie), Forschugszentrum Juelich Gmbh, \\
5170 Juelich, Germany \\
and\\
L.D.Landau Institute for Theoretical Physics, \\
GSP-1, 117940, ul.Kosygina 2, 117 334 Moscow, Russia
\vspace{1.0cm}\\}
{\Large
ABSTRACT
\medskip\\}

\end{center}
I present a new determination of the total cross section
for proton-proton collisions from the recent Akeno [1] results
on absorption of the cosmic ray protons in the
$p$-Air collisions. Extrapolation to
the SSC energy suggests
$\sigma_{tot}(p-p) \approx  (160-170)\, mb$.
I also comment on a possible sensitivity of the $p$-Air cross
section determinations to assumptions on the inelasticity
of nuclear collisions at high energy.
\pagebreak

At present, the accelerator experiments give the total proton-proton
cross section up to $\sqrt{s}=1.8\,TeV$, and the cosmic ray
experiments remain a unique source of information on
the proton interaction cross sections at higher energies. Recently,
the Akeno collaboration has presented the results on $p$-Air
interactions up to $\sqrt{s}=24 \, TeV$ [1]. The subject of this
short communication is a reanalysis of these data in terms of
the proton-proton cross section. My major conclusion, based on
the comprehensive analysis [2,3] of the relationship between
the proton-nucleus and the proton-proton cross sections is that
the underestimated values of $\sigma_{tot}(p-p)$ were inferred
in [1].

By the nature of cosmic ray experiments the quasilealstic ({\sl Qel})
scattering of primary protons on nuclei which retains the incident
hadron is unobservable.  Therefore,
the cross section, measured in the Akeno exeperiment,
must be identified with the {\sl absorption} cross section
given by
\arr
\sigma_{abs}(p-Air)=\sigma_{tot}(p-Air)-\sigma_{el}(p-Air)-
\sigma_{Qel}(p-Air)=  \nonumber \\
\int d^{2}\vec{b}\left\{1-
\left[1-{1\over A}\sigma_{in}(p-p)T(b)\right]^{2A} \right\} \approx
\int d^{2}\vec{b}\left\{1-
\exp\left[-\sigma_{in}(p-p)T(b)\right] \right\}
\label{eq:1}
\endarr
Here
$T(b)$ is the conventional optical thickness of a nucleus, folded
with the profile function $\Gamma_{pp}(b)$ of the $p-p$ scattering,
\beq
T(b)={2 \over \sigma_{tot}(p-p}\int d^{2}\vec{c}\Gamma_{pp}(\vec{c})
\int dz n_{A}(z,\vec{b}-\vec{c})\,\, ,
\label{eq:2}
\endeq
$n_{A}(\vec{r})$ is the nuclear matter density for the pointlike
nucleons (the usually quoted nuclear charge distributions include
the charge radius of the proton), and $\Gamma_{pp}(\vec{b})$
is defined by
\beq
f_{pp}(q)={ik \over 2\pi}\int d^{2}\vec{b}\Gamma_{pp}(b)
\exp(-i\vec{q}\cdot\vec{b}) \approx
{ik \over 4\pi}\sigma_{tot}(p-p)\exp\left(-{1\over 2}B_{pp}q^{2}\right)
\label{eq:3}
\endeq

The above simple Glauber model [4] formulae
must be corrected for the
Gribov's
inelastic shadowing [5], which was done in [2]. Because of
strong absorption this inelastic shadowing correction is
numerically small
(for the more details see [2,6]).

The major observation
is that by virtue of the $s$-channel unitarity there is a strong
correlation between the $p-p$ interaction radius and the $p-p$
cross section [2]. This correlation is less stringent than the one
dictated by the naive geometrical scaling
$B_{pp} \propto \sigma_{in}(p-p)$, but sufficiently strong to
ensure that there emerges a universal relationship
\beq
\sigma_{in}(p-p) =(100\, mb)[\sigma_{abs}(p-Air)/507\,mb]^{1.89}
\label{eq:4}
\endeq
The remarkable virtue of this relationship is that it does not
depend on the detailed form of the energy dependence of the
$p-p$ cross section. This relationship, derived in $[2]$,
includes corrections for Gribov's inelastic shadowing .
A very conservative estimate of the theoretical systematic
uncertainty of the inelastic $p-p$ cross section
determined from (\ref{eq:1}),(ref{eq:4}) does not exceed $5 \, mb$.
Gaisser et al.
[3] have studied in much detail $\sigma_{abs}(p-Air)$ as
a function of $B_{pp}$ and $\sigma_{in}(p-p),\, \sigma_{tot}(p-p)$
and have found results very close to (\ref{eq:4}).

The total cross section $\sigma_{tot}(p-p)$ is obtained from
(\ref{eq:4}) adding the elastic cross section:
$\sigma_{tot}(p-p)=
\sigma_{in}(p-p)+\sigma_{el}(p-p)$.
By virtue of the same $\sigma_{in}(p-p)-B_{pp}$ correlation,
the ratio
$
R=\sigma_{el}(p-p)/\sigma_{in}(p-p)
$
too is, to a crude
approximation, a function of predominantly the absolute
value of $\sigma_{in}(p-p)$. In the QCD motivated models of
the pomeron [2] the ratio $R$ rises from
$R \approx 0.19$ at $\sigma_{in}(p-p)=30 mb$ to
$R \approx 0.35$ at $\sigma_{in}(p-p)=100 mb$ and
$R \approx 0.46$ at $\sigma_{in}(p-p)=150 mb$.
Such a rise of $R$ is typical for the onset of the black
disc regime.  With the parameters
of the QCD pomeron which are consistent with the
$Sp\bar{p}S$ [7] and Fermilab [8] data,
the predictions [2] for $R$
in the vicinity of $\sigma_{in}(p-p)\sim 100\,mb$ follow
an approximate law
\beq
R\approx 0.1 + 0.25*(\sigma_{in}(p-p)/ 100\,mb)
\label{eq:5}
\endeq
For instance, E710 measured $R=0.23 \pm 0.012$ at
$\sigma_{in}(p-p)= 55.5 \pm 4 mb$ compared to
$R=0.24$ given by eq.(\ref{eq:5}).
The conservative estimate of the theoretical systematic
error in $R$ at $\sigma_{in}(p-p) \sim 100 mb$ is
$\Delta R \sim 0.05$.
The overall theoretical uncertainty in the so-determined
$\sigma_{tot}(p-p)$
will not exceed $(5-10)\,mb$.
The results of my reanalysis of the Akeno data are shown
in Table 1:

\begin{table}[h]
\center\begin{tabular}{|c|c|c|c|} \hline
$\log_{10}E$ &  $\sigma_{abs}(p-Air)$ & $\sigma_{in}(p-p)$&
$\sigma_{tot}(p-p)$\\
(GeV) & (mb) & (mb) & (mb) \\
\hline
 7.17-7.41 & $480 \pm 33$ & $90 \pm 12$ & $120 \pm 15$ \\
 7.41-7.65 & $500 \pm 38$ & $97 \pm 14$ & $130 \pm 18$ \\
 7.65-7.89 & $537 \pm 33$ & $111 \pm 13$ & $154 \pm 17$ \\
 7.89-8.13 & $507 \pm 61$ & $100 \pm 22$ & $135 \pm 29$ \\
 8.13-8.37 & $498 \pm 64$ & $97 \pm 24$  & $129 \pm 30$ \\
 8.37-8.61 & $550 \pm 72$ & $117 \pm 29$ & $162 \pm 38$ \\
\hline
\end{tabular}
\caption{Determination of the inelastic and total
cross section for $p-p$ scattering from the Akeno
data on the absorption cross section for $p$-Air collisions.}
\end{table}

The values of $\sigma_{tot}(p-p)$ presented in Table 1 are
by $\sim 35 mb$ larger than those cited in [1]. The origin
of this difference is as follows:
The Glauber formula for the inelastic cross section
$\sigma_{in}(p-Air)=\sigma_{tot}(p-Air)-\sigma_{el}(p-Air)$
which includes the unobservable quasielastic cross
section, reads
\arr
\sigma_{in}(p-Air)= ~~~~~~~~~~~~~~~~~~~~~~~~~~~~~~\nonumber\\
\int d^{2}\vec{b}\left\{1-
\left[1-{1\over 2A}\sigma_{tot}(p-p)T(b)\right]^{2A} \right\} \approx
\int d^{2}\vec{b}\left\{1-
\exp\left[-\sigma_{tot}(p-p)T(b)\right] \right\}
\label{eq:6}
\endarr
Notice the close similarity between the
 $\sigma_{in}(p-p)-\sigma_{abs}(p-Air)$ relationship,
eq.(1), and the  $\sigma_{tot}(p-p)-\sigma_{in}(p-Air)$
relationship, eq.(6).
(This is a particular case of the
universality of nuclear cross sections discussed in [6]).
Therefore, if one treats the cosmic
ray cross section as $\sigma_{in}(p-Air)$ (which for the above
discussed reasons is illegitimate), then $\sigma_{in}(p-p)$
given by eq.(4) will erroneously be taken for
$\sigma_{tot}(p-p)$.
Such a determination of $\sigma_{tot}(p-p)$ will be short
of the elastic scattering cross section $\sigma_{el}(p-p)$,
which is precisely a discrepancy between the results of
[1] and the present study. Indeed, the authors of Ref.1
have identified their measured cross section with
$\sigma_{in}(p-Air)$ calculated by Durand and Pi [9].

It is worthwhile to comment, that the quantity measured
in the Akeno experiment, like in all the air shower experiments,
is the observed mean free path for the development of air
showers. The relation of this quantity to the true absorption
length depends on the so-called inelasticity coefficient
$K_{in}$ -
the rate of transfer of energy from the projectile proton
to the secondary particles. The analysis [1] assumed, in fact, the
energy independent elasticity coefficient $K_{el}$ - a fraction
of projectile's energy carried away by the leading particles.

To a crude approximation, the energy dissipation rate is
controlled by a sort of the transport cross section
$\sigma_{abs}(p-Air)(1-K_{el}(p-Air))$. According
to an analysis [2], in the $p$-Air collisions the elasticity
coefficient decreases from $K_{el}(p-Air) \approx 0.4$ at
$\sigma_{abs}(p-Air) \sim 300 \, mb$ typical of the
accelerator (Fermilab) energies, down to
$K_{el}(p-Air) \approx 0.31$ at
$\sigma_{abs}(p-Air) \sim 400 mb$ and
$K_{el}(p-Air) \approx 0.23$ at
$\sigma_{abs}(p-Air) \sim 500 mb$.(Other models of the
scaling violations and their implications for the air
shower development are discussed by Gaisser et al. [10]).
Therefore, assuming the exact Feynman scaling of the
projectile fragmentation spectra, one will overestimate
$\sigma_{abs}(p-Air)$. Very crude estimate of the so
introduced "scaling-violation bias",
\beq
\Delta \sigma_{abs} \sim -\sigma_{abs}(Akeno)
{\Delta K_{el}(p-Air) \over 1-K_{el}(p-Air)}  \,\, ,
\label{eq:7}
\endeq
suggests that the absorption cross sections cited in [1]
could have been overestimated by $\sim (70-80)mb$. This
possible systematic bias exceeds the error bars quoted in [1].
If the Akeno results are corrected for such a bias, then
the relationship (\ref{eq:4}) suggests that
the proton-proton cross sections quoted in Table 1 must be
lowered too:
\beq
{\Delta \sigma_{in}(p-p) \over \sigma_{in}(p-p)} \approx
1.9
{\Delta \sigma_{abs}(p-Air) \over \sigma_{abs}(p-Air)} \approx
1.9{\Delta K_{el}(p-Air) \over 1-K_{el}(p-Air)}  \,\, ,
\label{eq:8}
\endeq
This amounts to a possible scaling-violation bias of
$\Delta\sigma_{tot}(p-p) \sim - 30 \,mb$.
More detailed analysis of the
scaling-violation bias requires modelling the development
of showers similar to what is being done by Gaisser et al.
[10], which goes beyond the scope of this paper.
\medskip\\
{\bf Conclusions:} Reanalysis of the Akeno data on
absorption of the cosmic ray protons in the Earth
atmosphere gives the proton-proton scattering cross
section by $\sim 30 mb$ larger than found in [1].
Extrapolation of the results of this reanalysis
to the design energy of SSC suggests
$\sigma_{tot}(\sqrt{s}=40\, TeV) \approx (160-170)\, mb$.
The potentially large scaling-violation bias is pointed out,
which suggests that the Akeno analysis [1] might have
overestimated the proton-air absorption cross sections.

{\bf Acknowledgement:} I am grateful to Tom Gaisser for
the encouraging correspondence.
\pagebreak

\end{document}